\pdfoutput=1

\documentclass[11pt]{article}

\usepackage{acl}

\usepackage{times}
\usepackage{latexsym}

\usepackage[T1]{fontenc}
\usepackage{booktabs}
\usepackage{fontawesome}

\usepackage[utf8]{inputenc}

\usepackage{microtype}

\usepackage{inconsolata}

\usepackage{graphicx}

\title{Cross-Cultural Differences in Mental Health Expressions on Social Media}

\author{Sunny Rai$^{*\dag}$, Khushi Shelat$^{*\dag}$, Devansh R Jain$^\diamond$, Kishen Sivabalan$^\dag$, Young Min Cho$^\dag$, \\ \bf Maitreyi Redkar$^\spadesuit$, Samindara Sawant$^\clubsuit$, Lyle H. Ungar$^\dag$ \& Sharath Chandra Guntuku$^\dag$   \\
$^\dag$University of Pennsylvania, $^\diamond$Carnegie Mellon University,\\ $^\spadesuit$Indian Institute of Technology, Bombay, $^\clubsuit$Disha Counseling Center  \\
\texttt{\{sunnyrai, sharathg\}@seas.upenn.edu} \\}

\begin{document}
\maketitle

\def\thefootnote{*}\footnotetext{These authors contributed equally to this work}
\begin{abstract}
 Culture moderates the way individuals perceive and express mental distress. Current understandings of mental health expressions on social media, however, are predominantly derived from WEIRD (Western, Educated, Industrialized, Rich, and Democratic) contexts. To address this gap, we examine mental health posts on Reddit made by individuals geolocated in India, to identify variations in social media language specific to the Indian context compared to users from Western nations. Our experiments reveal significant psychosocial variations in emotions and temporal orientation.  
 This study demonstrates the potential of social media platforms for identifying cross-cultural differences in mental health expressions (e.g. seeking advice in India vs seeking support by Western users). Significant linguistic variations in online mental health-related language emphasize the importance of developing precision-targeted interventions that are culturally appropriate.
\end{abstract}

\section{Introduction}

Over 197 million individuals in India are diagnosed with mental health disorders~\cite{sagar2020burden}, a disproportionate majority of whom do not receive mental healthcare \cite{singh2018closing}. 
Generative AI technologies can facilitate affordable and easily accessible mental health assessment and support, especially in under-resourced contexts such as India~\cite{stade2024large}. 

Mental disorders, however, manifest differently across cultures \cite{manson1995culture}. 
For example, AI models trained on Black individuals' language fail to detect depression in Black individuals  \cite{rai2024key}. Moreover, the frequency of language markers such as self-focus and negative evaluation indicative of depression do not increase with depression in Black individuals' language. AI systems lacking awareness of cross-cultural differences in mental health communications may lead to misdiagnosis and health inequity \cite{bailey2009major, lewis2005depression}. 

In this paper, we examine \textit{how mental health expressions  specific to Indian context vary compared to the rest of the world} (RoW). 
Previously, \citet{de2017gender} examined depression patterns in the online language of individuals from majority world countries including India. 
Another related work examined help-seeking patterns to inform social media platform design \cite{pendse2019cross}. This paper bridges two critical gaps from previous literature
. First, we focus on mental health expressions specific to India, the world's most populous country, by mining Reddit threads. Second, we collaborate with clinical psychologists practicing in India to validate the empirical findings, providing a culturally informed assessment of cross-cultural comparisons of mental health expressions. 


To inform the research on culturally competent mental health models \cite{sue1998search}, we adopt interpretable features that are comprehensible to stakeholders such as psychologists and policymakers, for modeling cross-cultural variations in mental health language.  We use psychosocial word categories (e.g., Linguistic Inquiry and Word Count (LIWC)) and topic modeling (word clusters derived using LDA) to examine variations in language correlated with depression \cite{aguirre2021gender, loveys2018cross, burkhardt2022comparative, rai2024key}. 
Using a blend of cross-disciplinary interpretable language features and machine learning models,  we address the overarching question, \textit{if and how the mental health expressions of Indian users on social media are different from the rest of the world (RoW)} 
by answering the following: 
\begin{itemize}
\item How do linguistic expressions in Reddit posts of individuals experiencing mental health challenges in India differ from individuals in the RoW?
\item How well do data-driven insights on mental health expressions align with the experience of clinical psychologists in India? 
\item Do the linguistic expressions of depression specific to India differ significantly to distinguish individuals from India compared to those from RoW?  
\end{itemize}


\section{Data}
Reddit offers a platform for individuals to share their mental health journey and seek support anonymously, thereby making it a rich source to understand the symptomatology and sequelae of mental health. India ranked 3rd in Reddit's website traffic globally after the US and the UK in 2024\footnote{\url{https://worldpopulationreview.com/country-rankings/reddit-users-by-country}}. 
Previously, Reddit posts have been used for identifying shifts to suicidal ideation~\cite{de2016discovering},  depression symptoms~\cite{liu2023detecting}, and the mental health expressions of immigrants~\cite{mittal2023language}, among others. 

\subsection{Subreddits: Mental Health vs Control}
We extracted $3,195,310$ posts and comments from mental health-related subreddits (See Appendix \ref{appendix:Subreddit_list} for list of subreddits) using the PushShift API~\cite{baumgartner2020pushshift}. The largest portion of users (36.1\%) were members of \textit{r/depression}. We queried subreddits external to the mental health subreddits to create a control group.
 We removed deleted usernames and null messages. We considered users who posted at least 500 words (excluding comments) to ensure sufficient linguistic richness in users' language for further analysis. We limit the scope of this study to Reddit posts to obtain expressions of experiences with mental health challenges rather than interactions (as in comments) with others' mental health challenges.

 \begin{figure}[t]
  \includegraphics[width=\linewidth]{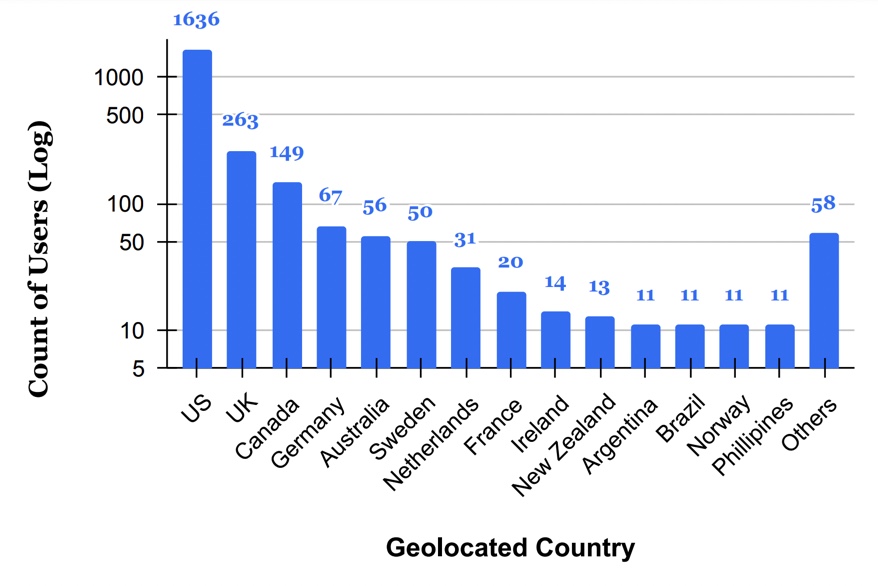}
  \caption{The count of users for each country in the Rest of World control group (log scale). Majority of users in the RoW group are geolocated to \textbf{Western countries}. The "Others" Category contains countries with less than 10 users, including Belgium (9), Italy (9),  Mexico (6), Malaysia (5), Romania (4), Croatia (4), UAE (2), South Africa (2), China (2), Spain (2), Greece (2), Denmark (1), Finland (1), Iceland (1), Japan (1), South Korea (1), Poland (1), Russia (1), Singapore (1), Thailand (1), Turkey (1) and Vietnam (1).}
  \label{fig:countrydist}
\end{figure}
 
 \subsection{Geolocation - India vs RoW} \label{sec:geolocation} 
 For labeling users' location, we performed two-level analysis. First, we identified India-focused subreddits (See Appendix \ref{appendix:Subreddit_list}) since users posting in India-specific subreddits are likely to be Indians and grouped together users into four groups: 
\begin{itemize}
    \item MH-India (4185 users): ``Individuals posting in India-specific Subreddits and posting in Mental Health Subreddits'', 
    \item MH-RoW (5588 users): ``Individuals NOT posting in India-specific Subreddits and posting in Mental Health Subreddits'',
    \item Control-India (2622 users): ``Individuals posting in India-specific Subreddits and NOT posting in mental health subreddits'' and,
    \item Control-RoW (5594 users): ``Individuals NOT posting in India-specific Subreddits and NOT posting in mental health subreddits''. 
\end{itemize}

The first group (MH-India) is our \textit{group of interest}; the remaining are controls.  
 
Second, we used the geolocation inference approach \cite{harrigian2018geocoding} as a second layer of verification for user location. The geolocation model is a location estimation model that utilizes word usage, the frequency distribution of subreddit submissions, and the temporal posting habits of each user to determine their location. Specifically, we use the pre-trained GLOBAL inference model\footnote{\url{https://github.com/kharrigian/smgeo/tree/master\#models}} to geolocate users in our dataset. 
We removed any users not geolocated to their group based on subreddit classification. For example, users in MH-India who are not geolocated to India and users in MH-RoW who are geolocated to India were removed. This functioned as a two-step verification to ensure that users in MH-India were from India. 

\paragraph{Manual Evaluation} We evaluated the quality of geolocation by manually verifying the self-disclosed location for randomly sampled 100 users. 
We found that the model's estimate of the individual's country matched the self-disclosed location, even though the state or city estimate was not always accurate.

Ultimately, 1200 users out of the initial 4185 users were left in the MH-India group, 
and 930 users out of 2622 were left in the Control-India group
. 
The majority of the users ($\approx 95\%$) in the RoW group were geolocated to Western nations (See Fig \ref{fig:countrydist}), affirming the dominance of West-centric data on Reddit. 
From now on, we use the terms RoW and Western nations interchangeably.


 \begin{figure}[t]
  \centering
  \includegraphics[width=0.9\linewidth]{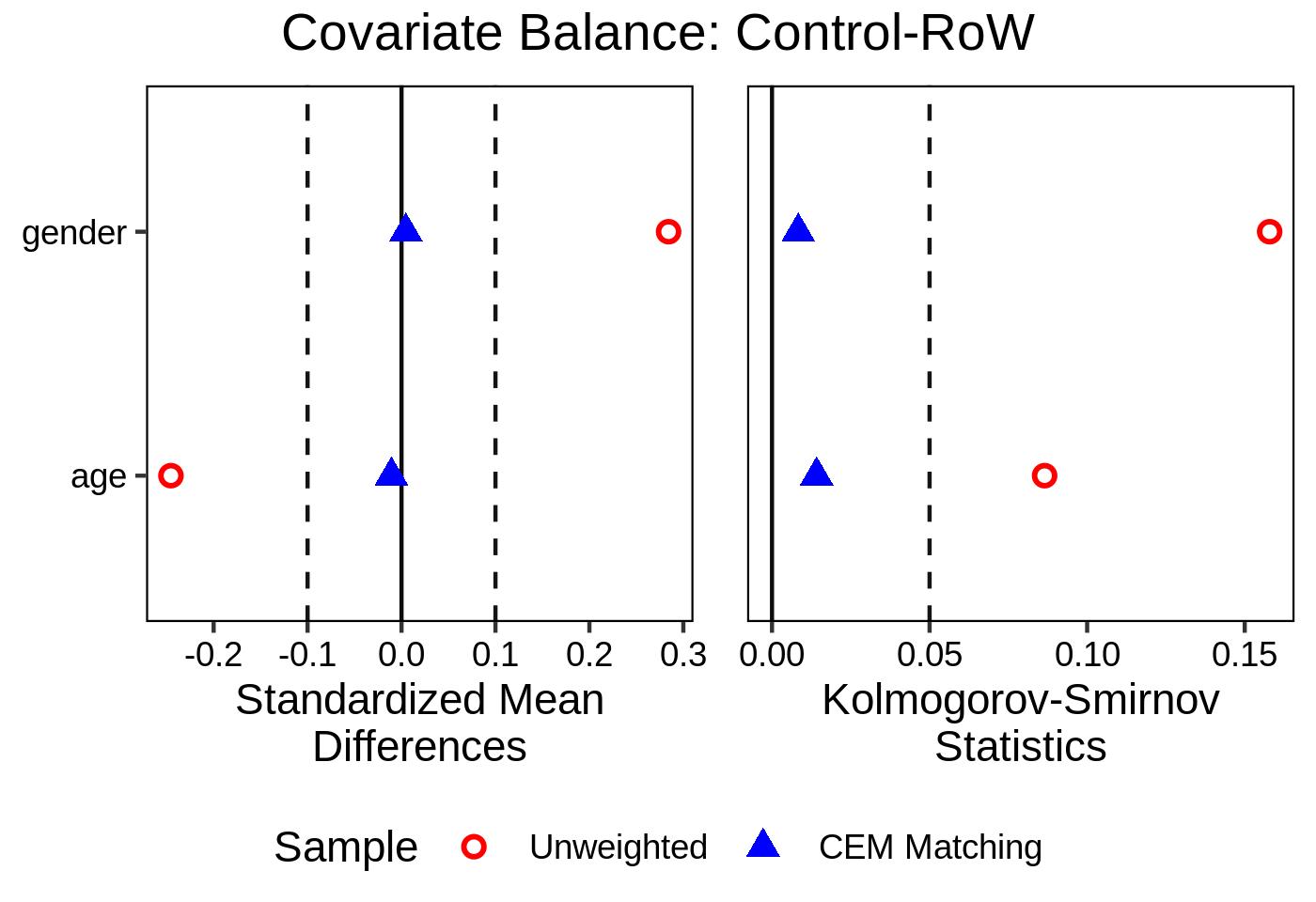} \label{fig:a}
  \includegraphics[width=0.9\linewidth]{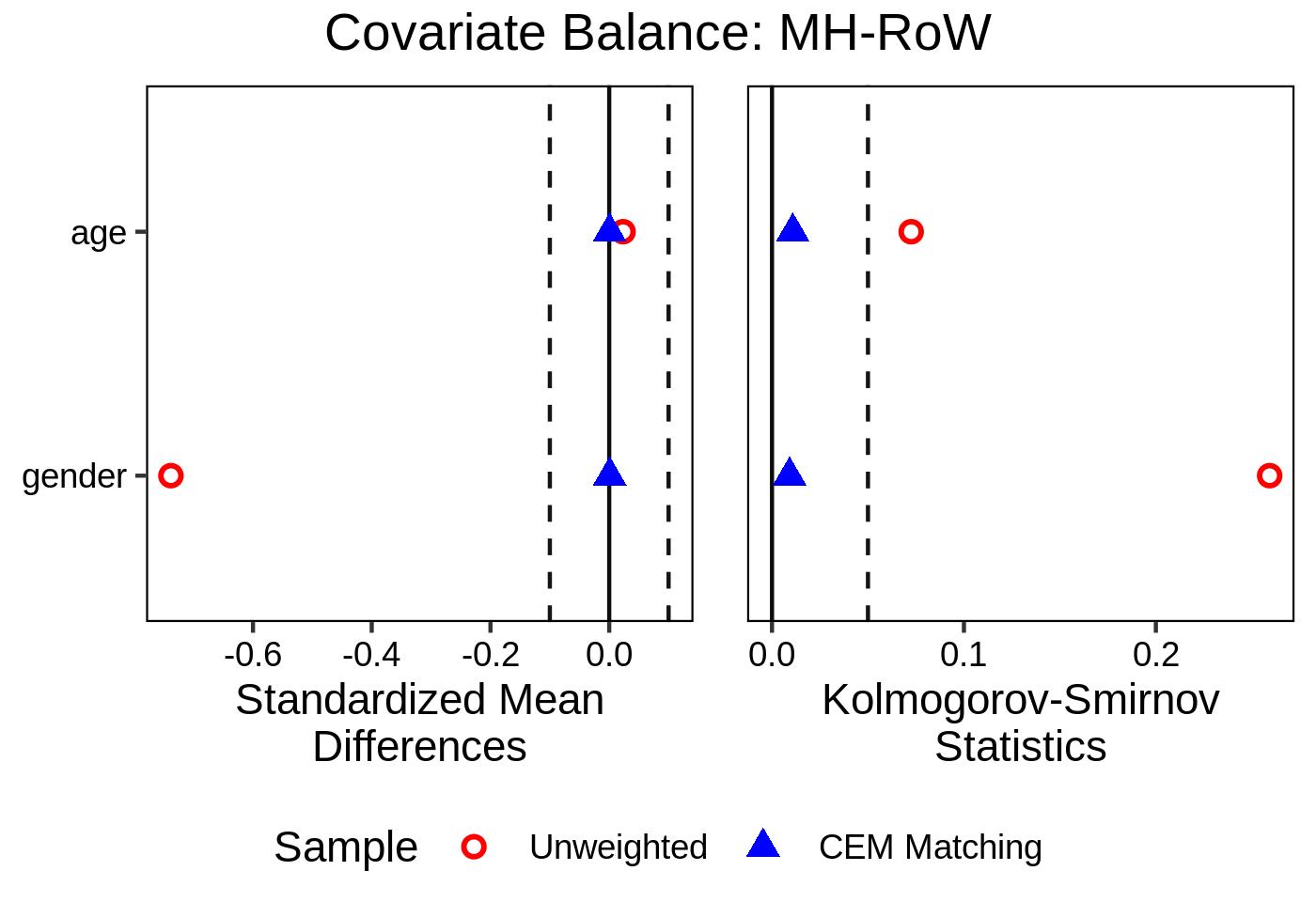}
  \caption{Differences in Covariates before and after CEM for groups ``Control-RoW'' and ``MH-RoW''. A control group is considered balanced with the treatment group if the difference is close to zero. Matching was not performed for Control-India due to smaller sample size.} \label{fig:AB}
\end{figure}

\subsection{Matching Control groups with users in MH-India} 

Age and gender are well-known confounders in behavioral health studies~\cite{schwartz2013personality}. We first estimated the age and gender of every user in our dataset using a machine-learning approach (see  Appendix \ref{appendix:ageGender} for method and evaluation). We then matched the users from our group of interest, i.e., MH-India, with those in control groups (MH-RoW, and Control-RoW) on these two covariates. Owing to the smaller sample size, matching was not performed for the Control-India group. However, the age distribution across the four groups was fairly similar before matching, with the average age being 25 for the MH-India, Control-India, and Control-RoW groups and 24 for the MH-RoW group. 

Ideally, the focus and control group samples should have indiscernible covariates. However, exact matching~\cite{rosenbaum2020modern} is difficult to achieve without dropping a large set of samples. Coarsened Exact Matching (CEM)~\cite{iacus2009cem} is a softer version of Exact Matching, which stretches the matching criteria wide enough to avoid dropping samples that are similar but not an exact match. 
We implement CEM using MatchIt package \cite{stuart2011matchit} in R and set the distance to ‘Mahalanobis’ for one-to-one matching. The quality of matching was evaluated using Standard Mean Differences and Kolmogorov-Smirnov Statistics (See Fig. \ref{fig:AB}). The mean age was 24.7 (sd= 3.41). The mean gender score was -0.97 (sd= 0.93), where a higher positive score indicates female. Table~\ref{tab:group_distribution} shows the total number of posts and users in each of the four groups after matching.

\begin{table}[!t]
\centering
\begin{tabular}{lrr}
\hline
\hline
 \textbf{Group} & \textbf{\# Distinct Users} & \textbf{\# Posts}   \\ \hline \hline
                
\textit{MH-India}   & 1200    & 50928   \\ 
\textit{Control-India}   & 930  & 69957\\
\textit{MH-RoW}    & 1200   & 54666   \\
\textit{Control-RoW} & 1200 & 122654  \\    \hline            
    Total      & 4530    & 298205    \\ \hline
\end{tabular}
   \caption{Number of users and posts in each of the four groups of our dataset.}
    \label{tab:group_distribution}
\end{table}

\section{Modeling Cross-Cultural Variations in Mental Health Language}


\subsection{Language Features}

We extracted a diverse set of interpretable language features widely used in Psychology literature to understand the unique markers of mental health among Indians.   
\begin{enumerate}
    \item  We extracted \textbf{1-3 grams} from posts and created a normalized bag-of-words representation for each user. We filtered out 1-3 grams having point-wise mutual information (PMI)$\leq$5. N-grams may reveal prevalent \textit{cultural idioms of distress} \cite{desai2017idioms} such as \textit{tension} \cite{weaver2017tension} that are unique to a culture.




\item {LIWC-22} is a closed dictionary comprising 102 \textbf{psychosocial categories}  to measure cognitive processes in language. These word 
categories in LIWC are counted for each user, and the count is normalized by the total number of 1-grams for each user, thereby representing each user as a vector of 102 normalized psychosocial categories.

\item We used {Latent Dirichlet Allocation} (LDA) \cite{blei2003latent} to extract latent \textbf{topics} in users' timeline data to capture themes behind mental distress specific to a culture. We do not use neural models or embeddings due to their skew toward West-centric data and inferences \cite{havaldar2023multilingual}.  

We generated three sets of topics by setting the number of topics $=[ 200, 500, 2000]$. 
We evaluated the topics' quality using Topic Uniqueness (TU)~\cite{nan2019topic}
. TU represents the number of times a set of keywords is repeated across topics; a higher TU corresponds to a rarely repeated word, indicating that topics are diverse, which is favorable.  Additionally, three co-authors independently reviewed the quality of topics. We set the number of topics to 2000 based on the automated 
and manual evaluation. 
Human annotators preferred high topic granularity because it highlights subtle linguistic variations across cultures.

 

\end{enumerate}

\subsection{Correlation Analysis}
To measure the association between language and the groups (i.e., MH-India vs MH-RoW), we performed ordinary least squares regressions with the feature sets. 
We calculated Pearson $r$ to measure the association of each feature to each group in a one-vs-all setting. 
p-values were corrected using Benjamini-Hochberg correction for multiple hypothesis testing. 
 23,344 1-3 grams, 102 word categories for LIWC, 2000 LDA topics, were considered for p-value correction.

\subsection{Predictive Model}
To examine whether the linguistic expressions of the MH-India group sufficiently differ from control groups, we trained `one vs rest' logistic regression models in a 10-fold cross-validation setting \cite{rifkin2004defense}. 
More sophisticated methods (such as XGBoost) could potentially provide higher performance, but the focus of the study is not to achieve state-of-the-art performance for group prediction but to test if sufficient discriminating evidence exists across groups to warrant culturally aligned models for estimating mental health risks. We report the Area Under the Receiver Operating Curves (AUC) for each feature for the MH-India and MH-RoW groups. 

\section{Results}

\begin{figure}[!t]
  \includegraphics[width=\linewidth]{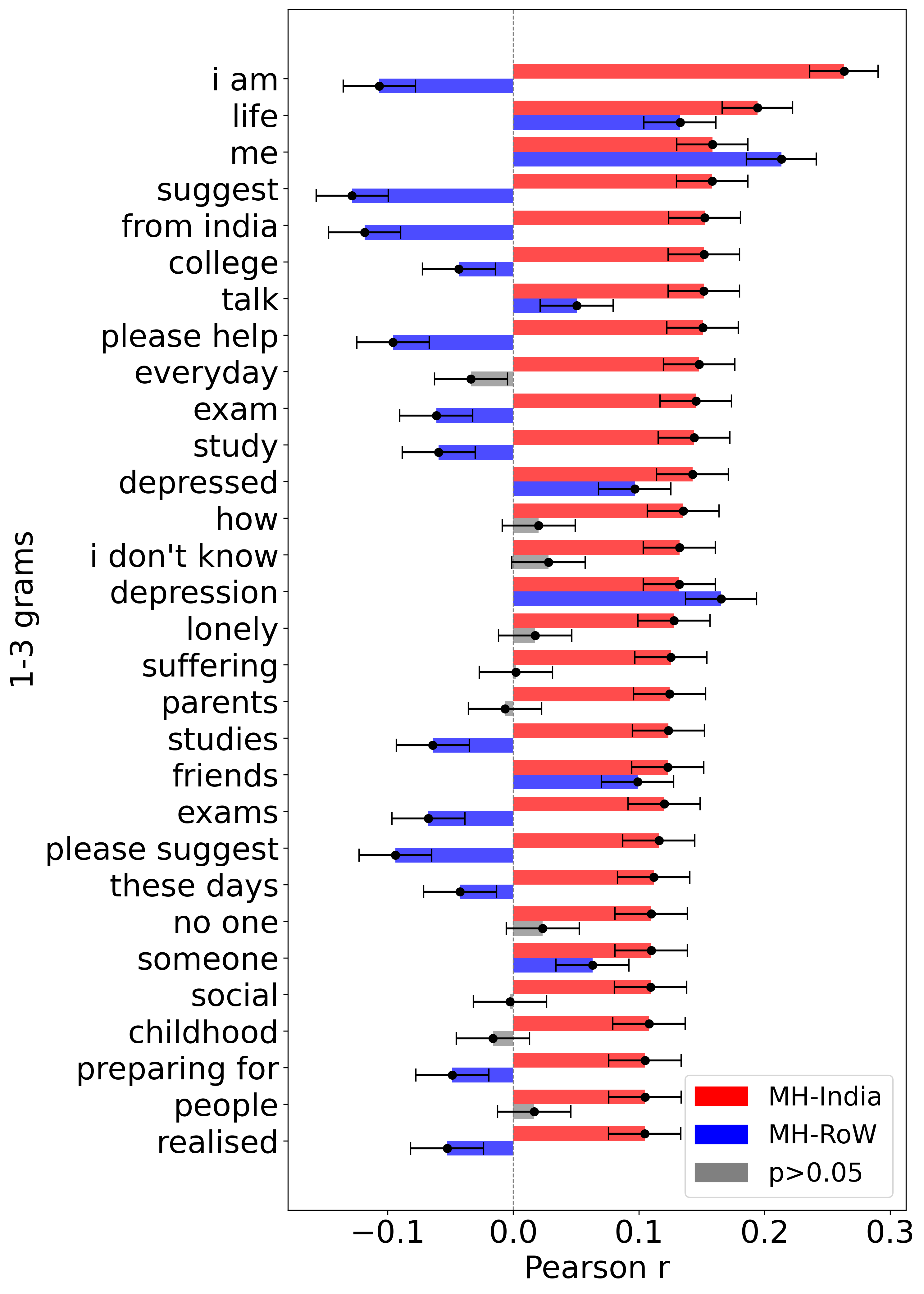}
  \caption{Top 30 statistically significant n-grams by effect size for MH-India and their corresponding correlation with MH-RoW. Significant at $p<.05$, two-tailed t-test, Benjamini-Hochberg corrected. Repetitive phrases (e.g. \textit{life} vs \textit{my life}) and function words are removed. See Fig \ref{fig:ngrams_India_RoW} for top n-grams for MH-RoW.}
  \label{fig:ngramResults}
\end{figure}

\subsection{Mental Health Expressions: India vs RoW}
\subsubsection{N-grams} 
Sixty-one 1-3 grams were significantly ($p<0.05$) correlated with the MH-India group, and 156 were correlated with the MH-RoW group. Figure~\ref{fig:ngramResults} illustrates the top 30 1-3 grams arranged in decreasing order of Pearson $r$ for MH-India. Introductory phrases (\textit{i am, from india}), negation (\textit{don't, don't know}), and help-seeking phrases (\textit{suggest, please help}) are exclusively correlated with MH-India, reflecting the struggle of users in MH-India for accessing mental help. Academic-related stressors (\textit{college, exam, study}) are exclusively seen in discussions in MH-India. This is particularly interesting, considering users in both groups were matched for age, yet discussions around student-life challenges are prevalent exclusively in MH-India. Self-referential pronouns such as \textit{I} and \textit{me}, negative feelings and symptoms mentions are more frequently seen in MH-RoW (see Fig. \ref{fig:ngrams_India_RoW}). 
Overall, the discussion in MH-India subreddits is centered around seeking help, whereas symptoms and diagnosis are more commonly discussed in the MH-RoW group.

\begin{figure}
    \centering
    \includegraphics[width=\linewidth]{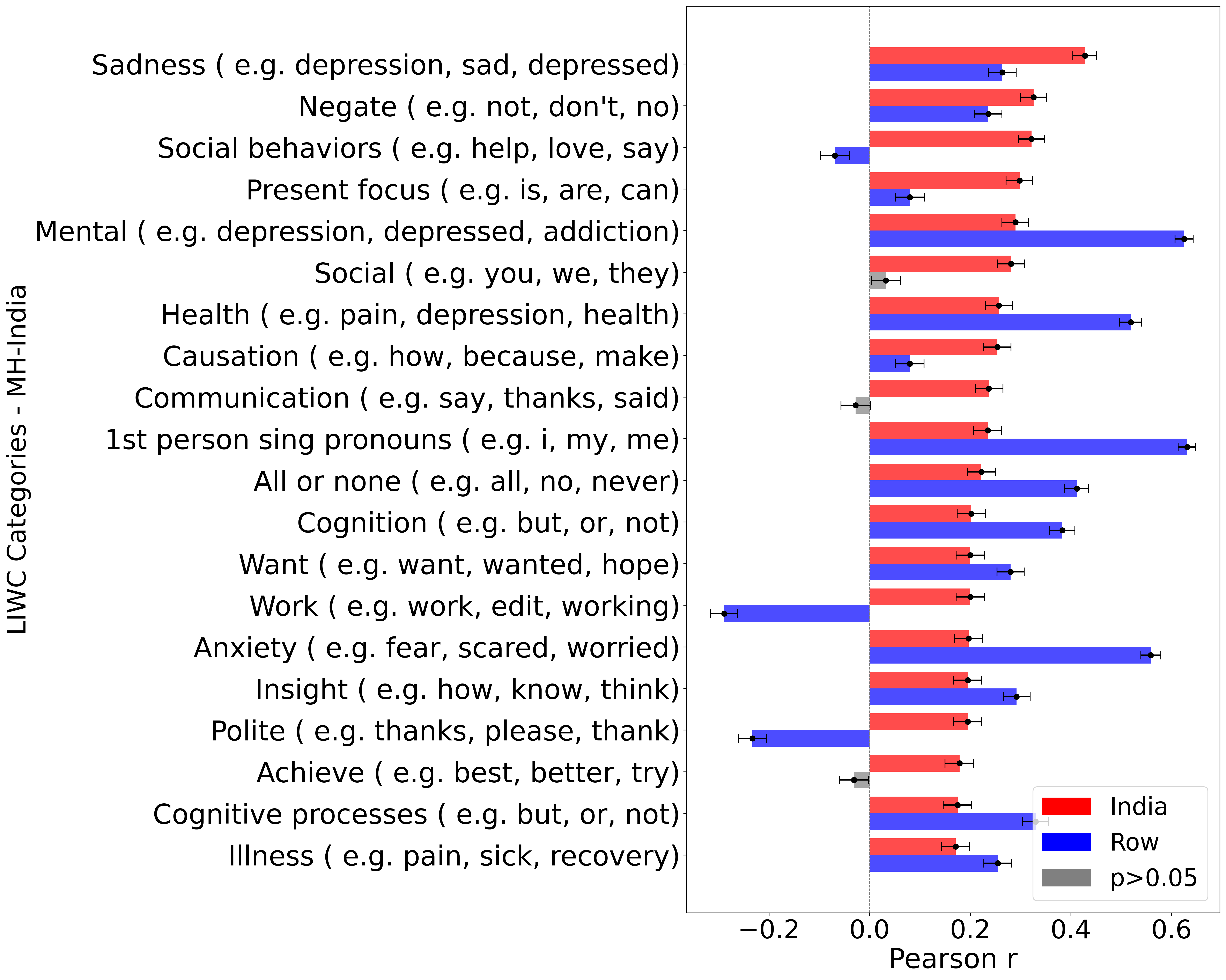}
    \caption{Pearson r for top 20 LIWC categories (with top-3 words) correlated with MH-India with corresponding correlation with MH-RoW. Bars in gray color indicate insignificant correlation ($p>0.05$). p-values were corrected using Benjamini-Hochberg correction. Sadness, social behaviors, and present focus are the most strongly correlated categories.}
    \label{fig:liwc_correlation_yindyd}
\end{figure}

\begin{figure}
    \centering
    \includegraphics[width=\linewidth]{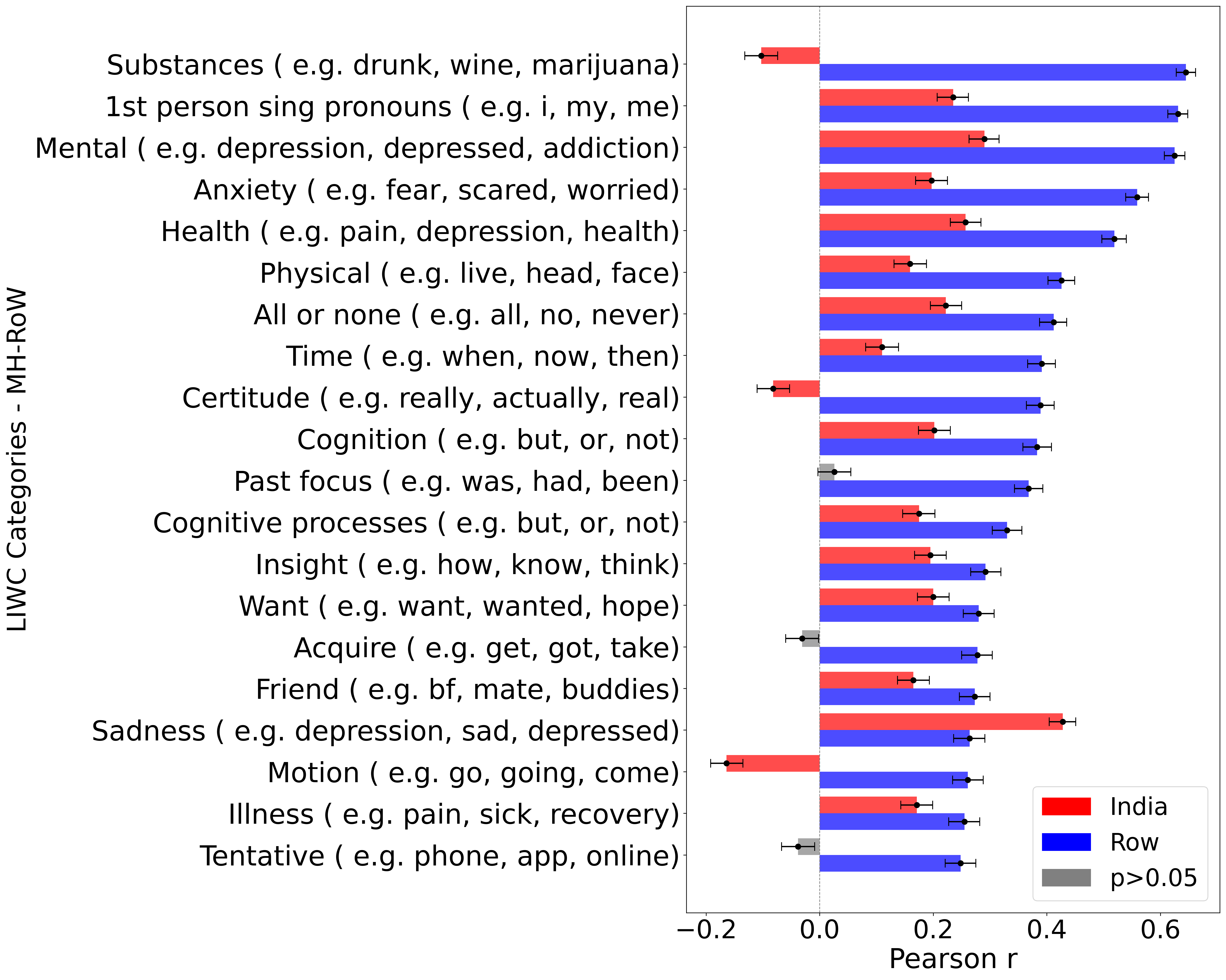}
    \caption{Pearson r for top 20 LIWC categories (with top-3 words) correlated with MH-RoW with corresponding correlation with MH-India. Bars in gray color indicate insignificant correlation ($p>0.05$). p-values were corrected using Benjamini-Hochberg correction. Substances, 1st person sing. pronouns, and Mental are the most strongly correlated categories.}
    \label{fig:liwc_correlation_nindyd}
\end{figure}

\subsubsection{LIWC} 
Fifty-two LIWC categories were significantly associated ($p<0.05$) with the MH-India group, whereas sixty categories were found to be correlated with the MH-RoW group. We provide the top LIWC categories for MH-India and MH-RoW in Fig \ref{fig:liwc_correlation_yindyd} and \ref{fig:liwc_correlation_nindyd}. Sadness, a subset of negative emotions is more strongly correlated with MH-India whereas anxiety (also a type of negative emotion) is more correlated with MH-RoW. Mental health-related discourse is more present-focused in India whereas in MH-RoW group, it is past-focused. This aligns with n-grams findings i.e. MH-India's emphasis on seeking help and MH-RoW's on discussing symptoms. Social aspects including the use of 2nd/3rd person pronouns (e.g. you, we),  communication, work, politeness, and achievement are positively correlated with only MH-India -- reflecting the potential sociocultural expectations (such as collectivism and \textit{high-power} distance in communication \cite{robert2000empowerment}) in Indian society. 

Discussions in MH-RoW use more clinical language with concepts from psychosocial categories such as mental, health, and illness (also observed in Fig. \ref{fig:ngrams_India_RoW}) - reflecting awareness about mental health disorders. Substances are the most strongly correlated psychosocial category with MH-RoW however, it has a negative correlation with MH-India. The correlation with 1st person singular pronouns is also three times more in MH-RoW compared to MH-India. This aligns with recent findings that the use of self-referential pronouns is more likely a marker of depression in European Americans compared to other demographic groups \cite{rai2024key}. Similarly, cognitive processes such as all or none (also a type of cognitive distortion \cite{mercan2023investigation}) and certitude are more strongly correlated with MH-RoW.  

Categories such as \textit{illness} and \textit{friend} are similarly correlated with both groups.

\begin{figure*}
    \centering
\includegraphics[width=0.8\linewidth]{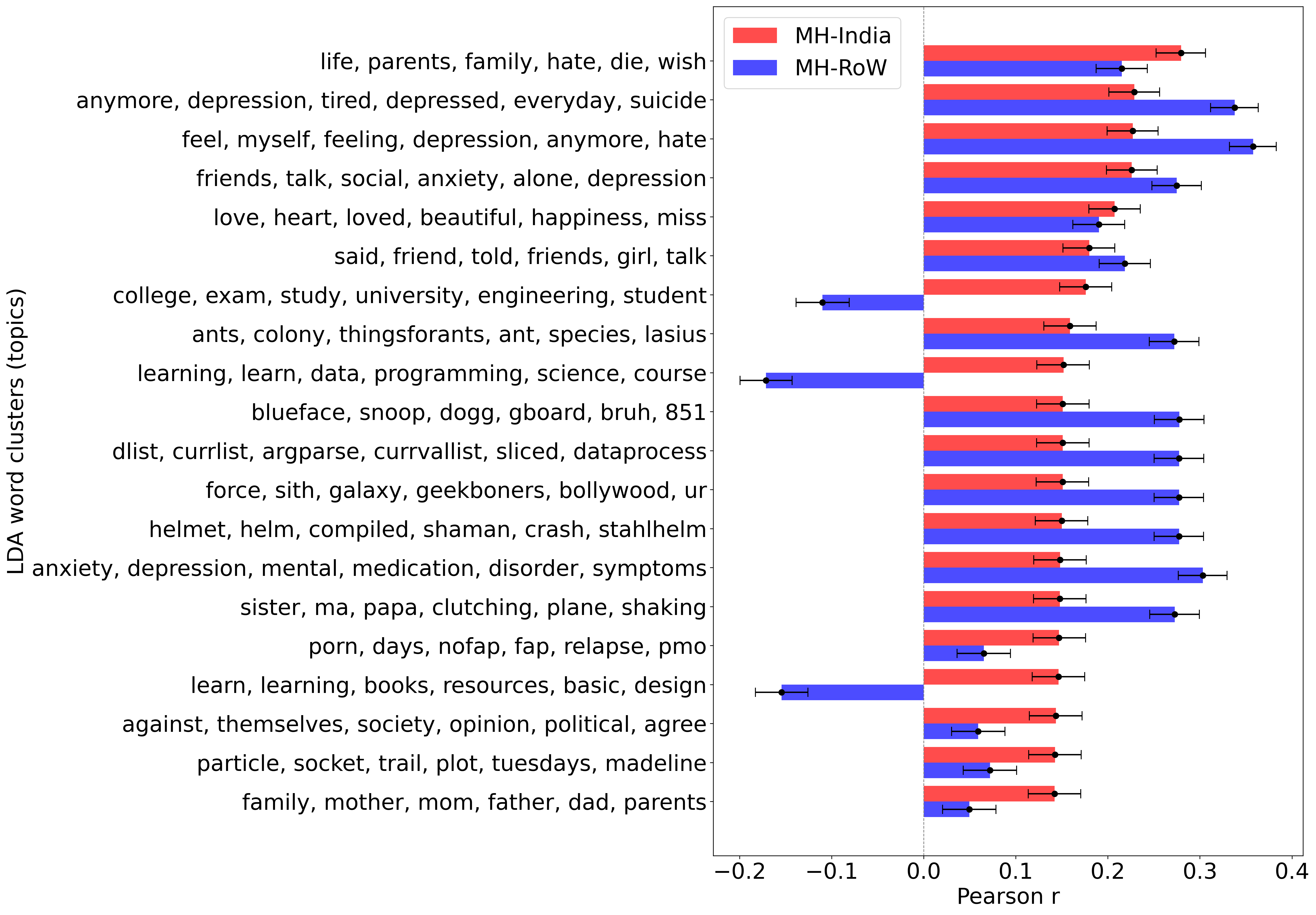}
    \caption{Top 20 LDA topics arranged in decreasing order of correlation with MH-India and its corresponding correlation with MH-RoW are shown. All topics shown are statistically significant at $p<.05$, two-tailed t-test, Benjamini-Hochberg corrected. }
    \label{fig:topics_output}
\end{figure*}

\subsubsection{LDA Topics} Of $2000$ topics, $109$ were found to be significant ($p < 0.05$) for the MH-India group and $216$ for MH-RoW group. The top topics and the corresponding Pearson \textit{r}, 
are provided in Fig. \ref{fig:topics_output}. Academic stressors  (e.g. \textit{college, exam, study, university}; \textit{learning, learn, data, programming}) are exclusively related to discussions in MH-India. Family (\textit{family, mother, mom, father}) and sexual content (\textit{porn, days, nofap, fap}) are more frequent in MH-India compared to MH-RoW. Other topics including negative/suicidal thoughts (\textit{life, parents, family, hate, die}), lack of belongingness (\textit{friends, talk, social, anxiety, alone}; \textit{said, friend, told, friends}), depression mentions (\textit{anymore, depression, tired, depressed}; \textit{anxiety, depression, mental, medication}) are overlapping however, they are more strongly correlated with MH-RoW. There are several topics (\textit{ants, colony, thingsforants, ant}; \textit{blueface, snoop, dogg, gboard}, etc,) which appear unrelated to mental health discussion but are discussed in both groups. 

\paragraph{Validation} We validated the significantly correlated topics for our group of interest (MH-India) by showing the top words 
to two clinical psychologists with significant practical experience with seeing patients and mental healthcare in India. Specifically, we asked the following question:

To what extent the open vocabulary topics having a significant correlation with the MH-India group are prevalent in Indian patients? - A Likert scale of 0-5 is provided where $5$ indicates `Highly Prevalent' and `0' indicates `Not observed at all'. 
    


\textbf{Prevalence} While independently labeling topics for prevalence, the clinical psychologists agreed with each other 
$81.49 \%$ of the time. Of the top 20 topics significantly associated with the MH-India group
, 95\% were ranked either extremely or somewhat prevalent (4 or 5 on a scale of 1 - 5) in India by at least one of the two clinical psychologists, and 80\% were ranked as prevalent (a score of 4 or 5) by both evaluators. 
Of the 109 topics significantly associated with the MH-India group, 56\% were annotated as prevalent by at least one evaluator. 

\subsection{Is the language in MH-India different from control groups? }

High AUC scores (See Table \ref{tab:prediction_results}) demonstrate that users' language in the MH-India group significantly differs from those in the control groups, including MH-RoW. 
All language feature groups (i.e., n-grams, LIWC, and LDA topics) have fairly high AUC, 
indicating differences in mental health expressions at 1-3 gm level, in psychosocial categories as well as in latent topics of discussion. 

\begin{table}[t]
    \centering
    \begin{tabular}{c|c c c}
    \hline
    \hline
         & 1-3 grams & LIWC & LDA Topics \\ \hline \hline
    MH-India   & 0.853   & 0.776 & 0.758 \\
    MH-RoW  & {0.881}  & 0.818 &  0.811\\
    \hline
    \end{tabular}
    \caption{AUCs for logistic regression one vs. rest models predicting group membership.}
    \label{tab:prediction_results}
\end{table}

\section{Discussion}
Our study uncovers significant cultural differences in the way users in the MH-India and MH-RoW groups discuss their mental health, underscoring the influence of sociocultural settings on how people perceive mental disorders and seek help. While MH-RoW discussions predominantly focus on feelings, symptoms, and peer support, MH-India situates their struggles with family, education, and work pressures, 
and are more likely to seek help or advice on social media platforms. \citet{pendse2019cross} also found that Indians discuss ``wanting or needing friends'' on Mental Health Support Forum more than other countries.

 Indian social media users often express immediate hopelessness or sadness with their current situation \cite{bahri2023language}. This differs from the in-person Indian patients having access to healthcare, who may ruminate more on past failures or anticipate future challenges.  
Similarly, somatization is widely observed among in-person patients in India \cite{kirmayer1998culture}, we, however, observe a weaker correlation for health-related mentions (such as pain, sick, etc.) with MH-India compared to MH-RoW. 
Social media data may underrepresent depression symptoms, as Indian users often seek advice rather than explicitly expressing distress. AI models trained on social media risk missing key depressive indicators, especially in cultures where emotional struggles are less directly verbalized. For instance, \citet{de2017gender} demonstrated that Indian and South Africa-based users are less candid in their posts and less likely to exhibit negative emotions in comparison to their Western counterparts.


Regarding latent themes in Reddit discussions, only $56\%$ of 109 LDA topics correlated with the MH-India group were labeled as \textit{prevalent} in Indian patients by clinical psychologists. 

Of Top-20 topics in MH-India which were labeled as "not prevalent" by psychologists comprise \textit{Video Games/Hobbies} and \textit{Programming/Learning}. This indicates the influence of digital content and growing isolation amongst the undiagnosed young population. These topics could be underrecognized concerns. 

The growing treatment gap for mental disorders is a major concern in Indian society. The economic loss from mental health conditions between 2012-2030 is estimated at USD 1.03 trillion\footnote{United Nations: \url{https://www.who.int/india/health-topics/mental-health}}. Automated systems that could diagnose and support mental well-being can potentially alleviate the lack of resources, but they would only be useful when designed considering the cultural sensitivities and norms of society. Based on our findings, culturally competent mental health intervention techniques should seek to bridge the gap to treatment by addressing hypochondriacal ideas and familial embarrassment in particular. This study establishes significant linguistic variations in the mental health-related language in social media posts by Indians compared to individuals from the rest of the world. 

\section{Background}


Depression and anxiety disorders are the most imminent mental health challenges, with the highest contribution to Indian Disability Adjusted Life Years 
~\cite{sagar2020burden}. 
Fear of shame is a  primary barrier to mental health recovery in India whereas, for example, substance abuse is the major hurdle in America
~\cite{biswas2016cross}. 
$71\%$ of Indians exhibited stigma when answering questions about mental health~\cite{live2018india}. 
Relatedly, somatic symptoms, hypochondriasis, anxiety, and agitation are more commonly seen in Indian patients compared to psychological symptoms 
~\cite{gada1982cross}. 
While the extent of mental health stigma and treatment (un)availability is often studied, it remains unknown how individuals suffering from mental disorders in India express and seek support on social media. There is accumulating evidence that suggests language markers of depression vary with demographics such as race \cite{rai2024key, aguirre2021qualitative}, immigrant status \cite{SharathImmigrantPaper}, and geographic location \cite{de2017gender}.  


\section*{Limitations}

The text-based geolocation of individuals in this study could potentially label Indians who later moved to other countries as Indians residing in India. Further, the Reddit user sample does not represent the general population, as evidenced by the mostly English language data in our India samples, although India has over 100 languages. In particular, we note that the majority of users were geolocated to Karnataka (a southern state in India) and that the age (ranging between 12 and 48) 
distributions are not necessarily representative. Our work shows the significant cultural themes observed in Indian society. However, Reddit posts represent a small population of India. While this analysis provides correlational insight into the data, it does not offer causal claims. 

\section*{Ethical Considerations}

While Reddit data is public, it may contain users' personal information, including city and town. We limited our analysis to country and state-level geolocation information to reduce the possibility of personally identifying individuals. Gender was predicted using a continuous scale, with extremes indicating masculinity and feminity. 
We exercised caution while presenting linguistic patterns and examples not to reveal any individual's timeline quotes.
Members of our team have not viewed or worked with individual-level granular data.  
When done ethically with respect to user anonymity and privacy, we believe this line of research could assist in understanding diverse individuals' mental health challenges and developing personalized interventions that improve the well-being and mental health of under-resourced communities.


\bibliography{custom}

\appendix
\setcounter{table}{0}
\setcounter{figure}{0}
\renewcommand{\thetable}{A\arabic{table}}
\renewcommand{\thefigure}{A\arabic{figure}}
\section{Subreddits Used to Extract the Raw Data} \label{appendix:Subreddit_list}
\paragraph{Mental Health Subreddits:} The mental health subreddit was obtained from prior works~\cite{sharma2018mental,saha2020understanding}. These include: r/Anxiety, r/bipolar, r/BipolarReddit, r/depression, r/sad, r/SuicideWatch, r/addiction, r/opiates, r/ForeverAlone, r/BPD, r/selfharm, r/StopSelfHarm, r/OpiatesRecovery, r/Sadness, r/schizophrenia, r/AdultSelfHarm

\paragraph{Control Subreddits:} All subreddits excluding Mental health subreddits.

\paragraph{India focused Subreddits: } r/india, r/mumbai, r/tamil,
       r/Hindi, r/Kerala, r/Urdu, r/delhi, r/pune, r/hyderabad,
       r/bangalore, r/kolkata, r/telugu, r/marathi,
       r/AskIndia, r/sanskrit, r/Kochi, r/Rajasthan, r/pali,
       r/Chandigarh, r/Chennai, r/karnataka, r/Bhopal,
       r/Coimbatore, r/kannada, r/TamilNadu, r/Trivandrum, r/gujarat,
       r/punjabi, r/Bengali, r/kolhapur,
       r/Vijaywada, r/Dehradun, r/sahitya, r/Uttarakhand,
       r/ahmedabad, r/bharat, r/nagpur, r/Agra, r/assam,
       r/Indore, r/surat, r/navimumbai, r/Goa, r/sikkim, r/lucknow,
       r/Bareilly, r/nashik, r/Allahabad, r/Durgapur, 
       r/Jamshedpur, r/Asansol,
       r/indianews, r/IndianGaming, r/IndiaSpeaks,
       r/indiameme, r/dankinindia, r/indiasocial

\section{Age and Gender} \label{appendix:ageGender}
We applied an open-source age and gender predictive lexica~\cite{sap2014developing} to obtain continuous values of age and gender. This lexicon was built over a set of over 70,000 users from social media and blogs and predicted age with a Pearson r of 0.86 and gender with an accuracy of 0.91 and has been applied reliably on Reddit data in prior studies~\cite{zirikly2019clpsych}. We used the probabilities from this model to denote the gender attribute of users in our data and did not consider gender as a binary category.  

To validate the machine-generated predictions of gender and age for users within the Reddit dataset, we looked for posts containing self-disclosures of gender and age per user. Examples of this include “(23F)” for a user who self identifies as a 23 year old female. We were able to identify 5,844 posts across 706 unique users (See Table \ref{table:user_counts} for distribution) who employed some form of gender self-identification, allowing us to measure the accuracy of provided gender predictions. Using this subset, the model’s gender prediction holds at 91.89\% (See Table \ref{table:gender_accuracy} for groupwise performance).

\begin{table}[ht]
\centering
\begin{tabular}{l|c|c}
\hline
\textbf{Group Name} & \textbf{Gender} & \textbf{Age} \\ \hline
MH India & 224  & 427\\ \hline
MH RoW & 195  & 378\\ \hline
Non-MH India & 45 & 111 \\ \hline
Non-MH RoW & 242 &   388\\ \hline
\end{tabular}
\caption{Distribution of users who self-disclosed gender and age by Group}
\label{table:user_counts}
\end{table}

\begin{table}[ht]
\centering
\begin{tabular}{l|c|c}
\hline
\textbf{Group Name} & \textbf{ Accuracy} & \textbf{MAE (in yrs)} \\ \hline
MH India & 87.67\% & 3.12 \\ \hline
MH RoW & 86.69\% & 4.65\\ \hline
Non-MH India & 94.33\% & 4.38 \\ \hline
Non-MH RoW & 96.12\% & 6.48 \\ \hline
\end{tabular}
\caption{Model performance for predicting gender and age. MAE stands for Mean Absolute Error and is reported in years.}
\label{table:gender_accuracy}
\end{table}

\begin{figure}[ht]
    \centering
    \includegraphics[width=\linewidth]{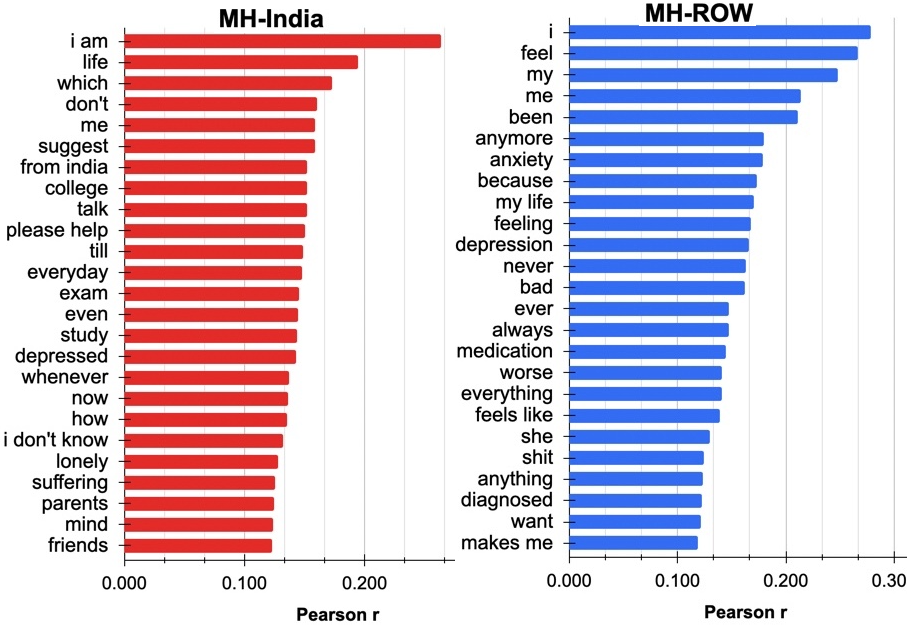}
    \caption{Top 25 1-3 grams in order of decreasing Pearson r for MH-India and MH-RoW.}
    \label{fig:ngrams_India_RoW}
\end{figure}

\section{Communication with Clinical Psychologists}

Table \ref{tab:annotation_guidelines} shows the email text used for communication with each clinical psychologist. 

\begin{table*}[t]
    \centering
       \caption{Email communication with clinical psychologists who performed an informed review of the topics. }
    \label{tab:annotation_guidelines}
    \begin{tabular}{p{14cm}}
    \toprule
     The goal of this project is to study the manifestation of mental illness in Indians. As a part of this project, we have identified a set of 100 Topics/ Themes that Indians Users were found to commonly discuss on Reddit, a social media platform. We have labeled these topics as per our understanding and we now need your help in interpreting these topics from your perspective. The objective is to essentially identify 
     
 Topic or theme of discussion in the context of mental illness in India, \\
How often a theme is observed in an Indian patient suffering from a mental illness? \\ 

These identified topics are available in this Google sheet. Please read the below steps carefully:\\ 
  Peruse the top words given in Column-A. These are the top 10 common words comprising a single topic.\\ 
   In Column E, select the degree of prevalence of this topic amongst Indian patients. The options are Highly prevalent, somewhat prevalent, unsure, rarely observed, and Not observed at all. \\ 
 You may add your comments in Column -F \\ 

     \bottomrule 
    \end{tabular}
 
\end{table*}

\end{document}